\pdfoutput=1
\documentclass[preprint,12pt]{elsarticle}

\usepackage[fleqn]{amsmath} 
\usepackage{amssymb} 
\usepackage{amsfonts} 
\usepackage{subfig}
\usepackage{geometry}
\usepackage{pdfpages}

\begin{document}
\begin{frontmatter}

\title{Analytic Solution of the Electromagnetic Eigenvalues Problem in a Cylindrical Resonator}

\author{M. Checchin\textsuperscript{a,b} and M. Martinello\textsuperscript{a,b}}

\address{\textsuperscript{a}Fermi National Accelerator Laboratory, Batavia, IL 60510\\
		 \textsuperscript{b}Illinois Institute of Technology, Chicago, IL 60616}

\begin{abstract}
Resonant accelerating cavities are key components in modern particles accelerating facilities. These take advantage of electromagnetic fields resonating at microwave frequencies to accelerate charged particles. Particles gain finite energy at each passage through a cavity if in phase with the resonating field, reaching energies even of the order of $TeV$ when a cascade of accelerating resonators are present.
In order to understand how a resonant accelerating cavity transfers energy to charged particles, it is important to determine how the electromagnetic modes are exited into such resonators. In this paper we present a complete analytical calculation of the resonating fields for a simple cylindrical-shaped cavity. 
\end{abstract}

\end{frontmatter}

\section{The Wave Equation}
\label{S:1}
\begin{figure}[t]
\begin{center}
\includegraphics[scale=0.4]{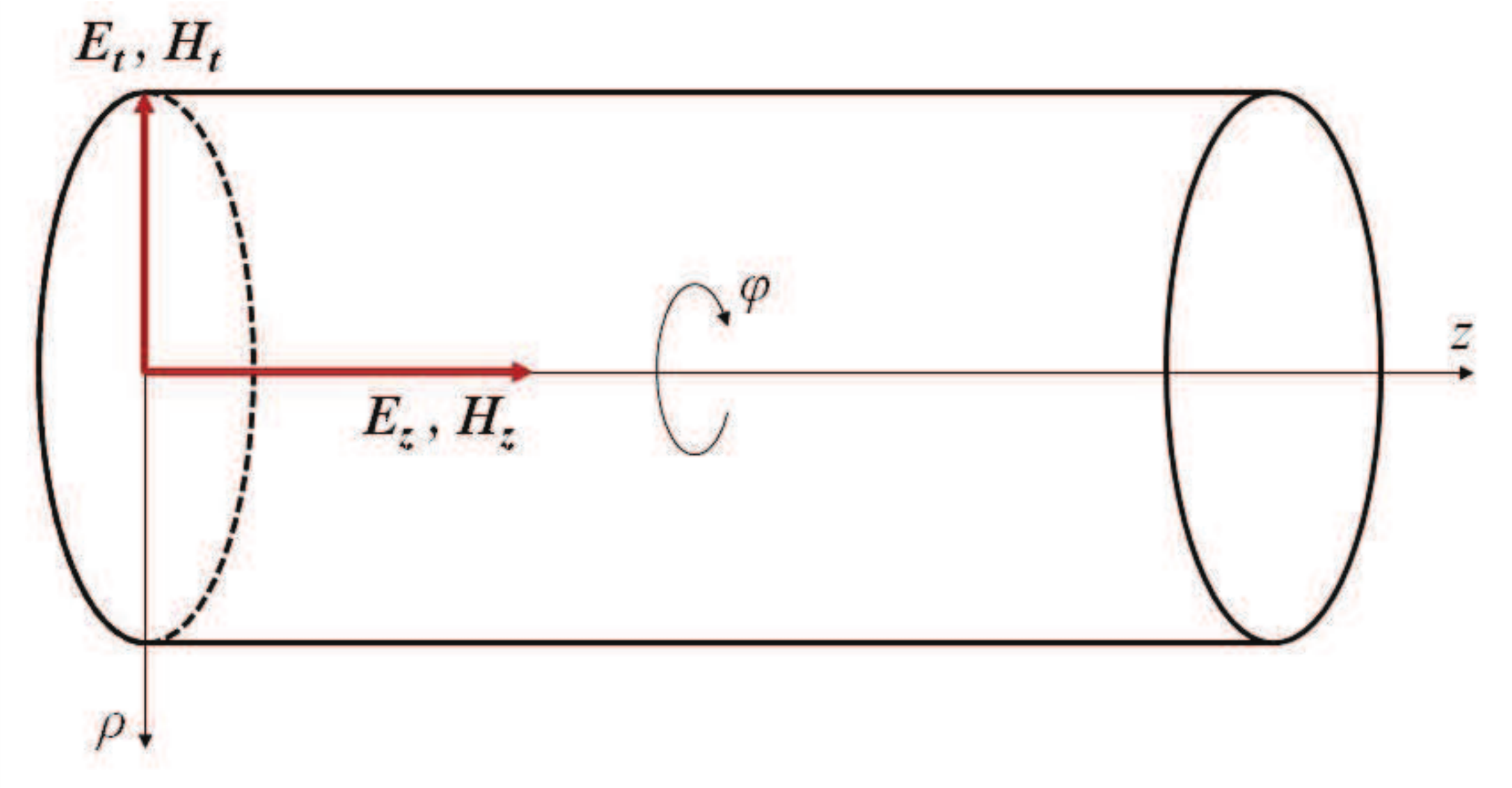}
\caption{Sketch of a perfect hollow cylindrical conductor, and of the internal decomposed field directions.}
\label{fig:Cyl}
\end{center}
\end{figure}

Let us take into account a hollow cylinder with perfect conducting walls and filled with vacuum (as shown in figure \ref{fig:Cyl}). The Maxwell equations inside the cylinder, where there are no charges $ \left( \rho =0\right)  $ and no currents $ \left(  \mathbf{J}=0 \right)$, are:
\begin{equation}
\begin{split}
\nabla\cdot\mathbf{E} = 0  \quad\quad & \nabla\times\mathbf{E} = -\mu_{0}\dfrac{\partial\mathbf{H}}{\partial t}\\
\nabla\cdot\mathbf{H} = 0  \quad\quad & \nabla\times\mathbf{H} = \varepsilon_{0}\dfrac{\partial \mathbf{E}}{\partial t}
\end{split}
\label{eq1}
\end{equation}
where $ \mathbf{E} $ is the electric field, $ \mathbf{H} $ is the magnetic field, $ \mu_{0} $ is magnetic permeability and $\varepsilon_{0}$ the magnetic permittivity of the vacuum.

Taking the curl of $\nabla\times\mathbf{E}$ and $\nabla\times\mathbf{H}$, after some arrangements the wave equation can be defined for both the electric end the magnetic field as:
\begin{equation}
\left( \nabla^2 - \dfrac{1}{c^2}\dfrac{\partial^2}{\partial t^2}\right) 
	\left\lbrace 
	\begin{array}{c}
		\mathbf{E} \\
		\mathbf{H}
	\end{array} 
	\right\rbrace = 0
\label{eqMax1}
\end{equation}

Defining the $z$-axis as the axis of the cylinder, it is convenient to select the spatial variation of the field in the $z$ direction, so equation \ref{eqMax1} can be rewritten as:
\begin{equation}
\left( \nabla^2 - \dfrac{1}{c^2}\dfrac{\partial^2}{\partial t^2}\right) 
	\left\lbrace 
	\begin{array}{c}
		\mathbf{E_{z}} \\
		\mathbf{H_{z}}
	\end{array} 
	\right\rbrace = 0
\label{eqMax2}
\end{equation}
where $c^{2}=1/\varepsilon_{0}\mu_{0}$ is the speed of light.

Since we are in a cylindrical symmetric system, a cylindrical system of coordinates is more indicate than a cartesian one to solve this problem, thus electric and magnetic fields will be defined as: $E_{z}=E_{z}(\rho,\varphi,z,t)$ and $H_{z}=H_{z}(\rho,\varphi,z,t)$.

The wave equation (equation \ref{eqMax2}) has the same form for electric and magnetic fields along the $ z $-direction, therefore we will solve it by replacing $E_{z}(\rho,\varphi,z,t)$ and $H_{z}(\rho,\varphi,z,t)$ with $u(\rho,\varphi,z,t)$. Therefore, considering the fictitious function $u(\rho,\varphi,z,t)$, the wave equation becomes:
\begin{equation}
\nabla^2 u-\dfrac{1}{c^2}\dfrac{\partial^2 u}{\partial t^2}=0
\label{eq:1.1}
\end{equation}
where, the Laplacian operator in cylindrical system of coordinates is:
\begin{equation}
\begin{split}
\nabla^2 &=\dfrac{\partial^2}{\partial \rho^2}+
			\dfrac{1}{\rho}\dfrac{\partial}{\partial \rho}+
			\dfrac{1}{\rho^2}\dfrac{\partial^2}{\partial \varphi^2}+
			\dfrac{\partial^2}{\partial z^2}\\
		&=\nabla_{t}^{2}+\dfrac{\partial^2}{\partial z^2}
\end{split}
\label{eq:1.2}
\end{equation}
Our goal is to completely define the field inside a cylindrical cavity in every its spatial components $\rho$, $\varphi$ and $z$. It is therefore useful to define the Laplacian as a linear combination of two differential linear operators, $\nabla_{t}^{2}$ and $\partial^2/\partial z^2$, in this way we can separate the problem in its longitudinal and transverse components.

It follow that the separation of variables solution we are looking for has the form:
\begin{equation}
u(\rho,\varphi,z,t)=Y(\rho,\varphi)Z(z)T(t)
\label{eq:YZT}
\end{equation}
where $Y(\rho,\varphi)=R(\rho)\Phi(\varphi)$.\\
So, substituting $u(\rho,\varphi,z,t)$ and dividing by $YZT$ we get:
\begin{equation}
\dfrac{1}{Y}\nabla_{t}^{2}Y+
\dfrac{1}{Z}\dfrac{\partial^2 Z}{\partial z^2}=
\dfrac{1}{Tc^2}\dfrac{\partial^2 T}{\partial t^2}
\label{eq:1.4}
\end{equation}
The temporal and longitudinal parts of the wave equation can be easily solved by applying the separation of variables methods, using $-\omega^2$ and $-\kappa^2$ as separation constants for the temporal and longitudinal equations respectively.

The temporal equation has the form:
\begin{equation*}
\dfrac{T''}{T}=-\omega^2 \quad $;$\quad T''+\omega^2 T=0   
\label{eq:1.6}
\end{equation*}
which general solution is:
\begin{equation}
T(t)=Acos(\omega t)+Bsin(\omega t)=Ce^{\pm i\omega t}$,$
\label{eq:1.7}
\end{equation}
The longitudinal equation is instead:
\begin{equation*}
\dfrac{Z''}{Z}=-\kappa^2 \quad $;$\quad Z''+\kappa^2 Z=0   
\label{eq:1.8}
\end{equation*}
with general solution:
\begin{equation}
Z(z)=Acos(\kappa z)+Bsin(\kappa z)$,$
\label{Zequat}
\end{equation}
The wave equation can be now rewritten as an eigenvalues equation with the form:
\begin{equation}
\nabla_{t}^{2}u+\beta^2u=0\\
\label{Helmholtz}
\end{equation}
such equation is called \textit{Helmholtz equation}, and $\beta^2=\left(\omega/c\right) ^2-\kappa^2$ are the eigenvalues.

\section{Field Components Decomposition}
\label{S:2}
With the purpose of completely define the field inside a cylindrical hollow conductor, it is mandatory to decompose the electric and magnetic fields into their longitudinal ($z$) and transversal ($\rho$, $\varphi$) components (see figure \ref{fig:Cyl}). In order to do so, let us define the $\nabla$ operator and the field in their components:
\begin{equation}
\begin{split}
& \nabla=\nabla_{t}+\hat{z}\dfrac{\partial}{\partial z}\\
& \mathbf{E}(\rho,\varphi,z,t)=\mathbf{E_{z}}(\rho,\varphi,z,t)+\mathbf{E_{t}}(\rho,\varphi,z,t)\\
& \mathbf{H}(\rho,\varphi,z,t)=\mathbf{H_{z}}(\rho,\varphi,z,t)+\mathbf{H_{t}}(\rho,\varphi,z,t)\\
\end{split}
\label{definitions}
\end{equation}

For simplicity, let us also single out the spatial variation of the field along the $z$-axis as a simple travelling wave with form:
\begin{equation}
	\left( 
	\begin{array}{c}
		\mathbf{E}(\rho,\varphi,z,t) \\
		\mathbf{H}(\rho,\varphi,z,t)
	\end{array} 
	\right)
	=
	\left( 
	\begin{array}{c}
		\mathbf{E}(\rho,\varphi) \\
		\mathbf{H}(\rho,\varphi)
	\end{array} 
	\right)\cdot e^{\pm i\kappa z-i\omega t}
\label{Waves}
\end{equation}

Solving the derivative respect the time, the Maxwell equations (equation \ref{eqMax1}) become:
\begin{equation}
\begin{split}
    \nabla\times
	\left( 
	\begin{array}{c}
		\mathbf{E}\\
		\mathbf{H}
	\end{array} 
	\right)
	&=
	-i\omega
	\left( 
	\begin{array}{c}
		-\mu\mathbf{H}\\
		\varepsilon\mathbf{E}
	\end{array}
	\right) \\
	\nabla\cdot
	\left( 
	\begin{array}{c}
		\mathbf{E}\\
		\mathbf{H}
	\end{array} 
	\right)
	&=0
\end{split}	 
\label{Max3}
\end{equation}

Therefore, applying the definitions in equation \ref{definitions} and substituting the derivative respect $z$ with $\pm i\kappa$ (see equation \ref{Waves}), the Maxwell equations becomes:
\begin{equation}
\begin{split}
   \left( \nabla_{t}\pm i\kappa\hat{z}\right) \times 
	\left( 
	\begin{array}{c}
		\mathbf{E_{z}}+\mathbf{E_{t}}\\
		\mathbf{H_{z}}+\mathbf{H_{t}}
	\end{array} 
	\right)
	&=
	-i\omega
	\left( 
	\begin{array}{c}
		-\mu \mathbf{H_{z}}+\mathbf{H_{t}}\\
		\varepsilon \mathbf{E_{z}}+\mathbf{E_{t}}
	\end{array}
	\right)\\
	\left( \nabla_{t} \pm i\kappa\hat{z}\right) \cdot
	\left( 
	\begin{array}{c}
		\mathbf{E_{z}}+\mathbf{E_{t}}\\
		\mathbf{H_{z}}+\mathbf{H_{t}}
	\end{array} 
	\right)
	&=0
\end{split}	 
\label{Max4}
\end{equation}

Recalling that: $\hat{z}\times\hat{\rho}=\hat{\varphi}$, $\hat{z}\times\hat{\varphi}=-\hat{\rho}$, and rearranging these expressions equating like components, we are able to define the Maxwell equations decomposed into their transverse and longitudinal parts:
\begin{equation}
\begin{split}
    \nabla_{t}
	\left( \begin{array}{c}
		E_{z}\\
		H_{z}
	\end{array}\right)
	\times \hat{z}
	\pm i\kappa\hat{z}\times
	\left( \begin{array}{c}
		\mathbf{E_{t}}\\
		\mathbf{H_{t}}
	\end{array}\right)
	&=
	-i\omega	
	\left( \begin{array}{c}
		-\mu \mathbf{H_{t}}\\
		\varepsilon \mathbf{E_{t}}
	\end{array}\right)\\
	\nabla_{t}\times
	\left( \begin{array}{c}
		\mathbf{E_{t}}\\
		\mathbf{H_{t}}
	\end{array}\right)
	-i\omega
	\left( \begin{array}{c}
		\mu \mathbf{H_{z}}\\
		-\varepsilon \mathbf{E_{z}}
	\end{array}\right)
	&=0\\
	\nabla_{t}\cdot
	\left( \begin{array}{c}
		\mathbf{E_{t}}\\
		\mathbf{H_{t}}
	\end{array}\right)
	\pm i\kappa
	\left( \begin{array}{c}
		\mathbf{E_{z}}\\
		\mathbf{H_{z}}
	\end{array}\right)
	&=0
\end{split}	 
\label{Max5}
\end{equation}

Taking into account the first system of expressions in equation \ref{Max5}, the upper equation can be rewritten as:
\begin{equation*}
\begin{split}
\nabla_{t}E_{z}\times\hat{z}\pm i\kappa\hat{z}\times\mathbf{E_{t}}&=i\omega\mu\mathbf{H_{t}}\\
\hat{z}\times\left( \nabla_{t}E_{z}\times\hat{z}\right) \pm i\kappa\hat{z}\times\left(\hat{z}\times\mathbf{E_{t}}\right) &=i\omega\mu\hat{z}\times\mathbf{H_{t}}\\
\nabla_{t}E_{z}\mp i\kappa\mathbf{E_{t}}&=i\omega\mu\hat{z}\times \mathbf{H_{t}}
\end{split}
\end{equation*}
while, the lower one becomes:
\begin{equation*}
\begin{split}
\nabla_{t}H_{z}\times\hat{z}\pm i\kappa\hat{z}\times\mathbf{H_{t}}&=-i\omega\varepsilon\mathbf{E_{t}}\\
\hat{z}\times\left( \nabla_{t}H_{z}\times\hat{z}\right) \pm i\kappa\hat{z}\times\left(\hat{z}\times\mathbf{H_{t}}\right) &=-i\omega\varepsilon\hat{z}\times\mathbf{E_{t}}\\
\nabla_{t}H_{z}\mp i\kappa\mathbf{H_{t}}&=-i\omega\varepsilon\hat{z}\times \mathbf{E_{t}}
\end{split}
\end{equation*}

These two equations define a system with two unknown variables, $E_{t}$ and $H_{t}$:
\begin{equation}
\left\lbrace \begin{array}{c}
		\pm i\kappa\mathbf{E_{t}}=\nabla_{t}E_{z}-i\omega\mu\hat{z}\times\mathbf{H_{t}}\\
		\pm i\kappa\mathbf{H_{t}}=\nabla_{t}H_{z}+i\omega\varepsilon\hat{z}\times\mathbf{E_{t}}
	\end{array}\right. 
\end{equation}
so, the system can be solved as follow:
\begin{equation*}
\begin{split}
		i\kappa\mathbf{E_{t}}
		&=\pm\nabla_{t}E_{z}-
		\dfrac{i\omega^2\varepsilon\mu}{\kappa}\left( \hat{z}\times\left( \hat{z}\times\mathbf{E_{t}}\right) \right)-
		\dfrac{\omega\mu}{\kappa}\hat{z}\times\nabla_{t}H_{z} \\
		&=\pm\nabla_{t}E_{z}+
		\dfrac{i\omega^2\varepsilon\mu}{\kappa}\mathbf{E_{t}}-
		\dfrac{\omega\mu}{\kappa}\hat{z}\times\nabla_{t}H_{z} \\
		i\kappa\mathbf{H_{t}}
		&=\pm\nabla_{t}H_{z}-
		\dfrac{i\omega^2\varepsilon\mu}{\kappa}\left( \hat{z}\times\left( \hat{z}\times\mathbf{H_{t}}\right) \right)+
		\dfrac{\omega\mu}{\kappa}\hat{z}\times\nabla_{t}E_{z} \\
		&=\pm\nabla_{t}H_{z}+
		\dfrac{i\omega^2\varepsilon\mu}{\kappa}\mathbf{H_{t}}+
		\dfrac{\omega\mu}{\kappa}\hat{z}\times\nabla_{t}E_{z}
\end{split} 
\end{equation*}
Rearranging the two equations we get:
\begin{equation}
\begin{split}
	\mathbf{E_{t}}&=\dfrac{i}{\beta^2}\left( \pm\kappa\nabla_{t}E_{z}-\omega\mu\hat{z}\times\nabla_{t}H_{z}\right) \\
	\mathbf{H_{t}}&=\dfrac{i}{\beta^2}\left( \pm\kappa\nabla_{t}H_{z}+\omega\mu\hat{z}\times\nabla_{t}E_{z}\right)
\end{split} 
\label{Transverse}
\end{equation}

At this point, two class of electromagnetic fields inside a cylindrical hollow conductor can be discriminated: transverse magnetic ($TM$), and transverse electric ($TE$). $TM$ waves have only the electric field along the direction of propagation, while the $TE$ ones have only the magnetic field along the direction of propagation. Then, equation \ref{Transverse} should be rewritten separately for the two cases:
\begin{itemize}
\item Transverse Magnetic $H_{z}=0$:
\begin{equation}
\begin{split}
	\mathbf{E_{t}}&=\pm\dfrac{i\kappa}{\beta^2}\nabla_{t}E_{z}\\
	\mathbf{H_{t}}&=\dfrac{i\omega\mu}{\beta^2}\hat{z}\times\nabla_{t}E_{z}
\end{split}
\label{TM}
\end{equation}
\end{itemize}
\begin{itemize}
\item Transverse Electric $E_{z}=0$:
\begin{equation}
\begin{split}
	\mathbf{E_{t}}&=-\dfrac{i\omega\mu}{\beta^2}\hat{z}\times\nabla_{t}H_{z}\\
	\mathbf{H_{t}}&=\pm\dfrac{i\kappa}{\beta^2}\nabla_{t}H_{z}
\end{split}
\label{TE}
\end{equation}
\end{itemize}
Therefore, once the fields components along the $z$ direction ($E_{z}$ and $H_{z}$) are known, the other components for $TE$ and $TM$ waves can be calculated by using these equations.

\section{Solution of the Eigenvalues Problem}
\label{S:3}
In order to calculate the variation of the field inside the hollow conductor along the $z$-axis, let us consider again the Helmholtz equation (equation \ref{Helmholtz}), which corresponds to the eigenvalue partial differential equation of this system:
\begin{equation}
\nabla_{t}^{2}u+\beta^2u=0
\end{equation}
where $\beta^2=\left(\omega/c\right) ^2-\kappa^2$ are the eigenvalues, $\nabla_{t}^{2}=\left( \partial^2/\partial\rho^2\right)+\left( 1/\rho\right) \left( \partial/\partial\rho\right)+\left( 1/\rho^2\right) \left( \partial^2/\partial\varphi^2\right) $ is the transversal Laplacian operator and $u(\rho,\varphi,z,t)=R(\rho)\Phi(\phi)$ $Z(z)T(t)$ is the fictitious function that could correspond either to $E_{z}$ or $H_{z}$.

The Helmholtz equation can be rewritten as:
\begin{equation}
\left( \dfrac{\partial^2}{\partial \rho^2}+
			\dfrac{1}{\rho}\dfrac{\partial}{\partial \rho}+
			\dfrac{1}{\rho^2}\dfrac{\partial^2}{\partial \varphi^2}\right) u +\beta^2u=0
\end{equation}
so, substituting the solution $u(\rho,\varphi,z,t)$ and dividing by $R\Phi ZT$ we get:
\begin{equation}
\dfrac{1}{R}\dfrac{\partial^2 R}{\partial \rho^2}+
\dfrac{1}{R\rho}\dfrac{\partial R}{\partial \rho}+
\dfrac{1}{\Phi\rho^2}\dfrac{\partial^2 \Phi}{\partial \varphi^2}+
\beta^2=0
\end{equation}
in more compact notation:
\begin{equation}
\dfrac{R''+\dfrac{1}{\rho}R'}{R}+
\dfrac{1}{\rho^2}\dfrac{\Phi''}{\Phi}+
\beta^2=0
\label{eq:1.5}
\end{equation}
The separation of variables method is the applied, by choosing $-\nu^2$ as separation constant for the angular part:
\begin{equation*}
\dfrac{\Phi''}{\Phi}=-\nu^2 \quad $;$\quad \Phi''+\nu^2 \Phi=0   
\label{eq:1.10}
\end{equation*}
with general solution of the form:
\begin{equation}
\Phi(\varphi)=cos(\nu\varphi)+sin(\nu\varphi)=e^{-i\nu\varphi}+e^{i\nu\varphi}
\label{eq:1.11}
\end{equation}

This solution should respect the periodic boundary condition given by the cylindrical symmetry of the problem, $\Phi(\varphi)$ must return the same under a complete rotation of $\varphi$ around the $z$-axis: $\Phi(\varphi +2\pi)=\Phi(\varphi)$. Therefore:
\begin{equation*}
\begin{split}
\Phi(\varphi +2\pi)&= cos(\nu(\varphi +2\pi))+sin(\nu(\varphi +2\pi))\\
				&= e^{-i\nu(\varphi +2\pi))}+e^{i\nu(\varphi +2\pi))}\\
				&= e^{-i\nu\varphi}e^{-i2\nu\pi}+e^{i\nu\varphi}e^{i2\nu\pi}=e^{-i\nu\varphi}+e^{i\nu\varphi}
\end{split}
\end{equation*}
It means that $e^{\pm i2\nu\pi}=1$, so $\nu$ has to be a positive integer: $\nu =n=0,1,2...$. Then, the solution for the angular part will be:
\begin{equation}
\Phi_{n}(\varphi)=cos(n \varphi)+sin(n \varphi)=C e^{\pm n i\varphi}
\label{phisolution}
\end{equation}

Rewriting the partial differential equation (equation \ref{eq:1.5}) in terms of separation of variables we get the radial equation:
\begin{equation}
R''+\dfrac{1}{\rho}R'+
\left(\beta^2-\dfrac{n^2}{\rho^2}\right)R=0
\label{BesselEqBeta}
\end{equation}
The solutions of this equation could be \textit{standard Bessel functions} if $\beta >0$ or \textit{modified Bessel functions} if $\beta <0$.

Making the substitution $x=\beta\rho$, equation \ref{BesselEqBeta} becomes a \textit{Bessel equation}:
 \begin{equation}
R''+\dfrac{1}{x}R'+
\left(1-\dfrac{n^2}{x^2}\right)R=0
\label{BesselEq}
\end{equation}
This second order ordinary differential equation (ODE) can be solved by using the Frobenius method, the singularity at $\rho =0$ (or $ x=0 $) is indeed regular. Considering the most general form of second order ODE ($y''(x)+P(x)y'(x)+Q(x)y(x)=0$), we have:
\begin{equation*}
\begin{split}
P(x)=\dfrac{1}{x} \quad &\to\quad p_{0}=\lim_{x_{0}\to\infty}P(x)(x -x_{0})=1\\
Q(x)=1-\dfrac{n^2}{x^2} \quad &\to\quad q_{0}=\lim_{x_{0}\to\infty}Q(x)(x -x_{0})^2=-n^2
\end{split}
\end{equation*}

The indicial equation to calculate the roots $r_{1}$ and $r_{2}$ is then defined as:
\begin{equation*}
\begin{split}
& r(r-1)+p_{0}r+q_{0}=0\\
& r(r-1)+r-n^2=0\\
& r^2-n^2=0 \quad\to\quad r_{1,2}=\pm n
\end{split}
\end{equation*}

In this case $r_{2}-r_{1}=2n$ (where $n$ is an integer) and we are dealing with the second case of the Frobenius method and, likely, only one of the two solutions will be of the form of a simple infinite series $R_{1}(x)=\sum_{i=0}^{\infty}a_{i}x^{i+r_{1}}$. The second solution will be of the form: $R_{2}(x)=CR_{1}(x)ln(x)+\sum_{i=0}^{\infty}b_{i}x^{i+r_{2}}$ where $C=0,1$.

Considering the first root $r_{1}=n$ and plugging the generalized infinite series solution $R_{1}(x)=\sum_{i=0}^{\infty}a_{i}x^{i+n}$ in the Bessel equation (eq. \ref{BesselEq}) we get:
\begin{equation*}
\begin{split}
& \sum_{i} a_{i}(i+n-1)(i+n)x^{i+n-2}+
	\sum_{i} a_{i}(i+n)x^{i+n-2}+\\
&\quad\quad\quad\quad\quad\quad\quad\quad\quad\quad+\sum_{i} a_{i}x^{i+n}-
	\sum_{i} a_{i}n^2x^{i+n-2}
	=0\\
& \sum_{i} a_{i}\left[(i+n-1)(i+n)+(i+n)-n^2)\right]x^{i+n-2}+
	\sum_{i} a_{i}x^{i+n}
	=0\\
& \sum_{i} a_{i}\left[(i+n)^2-n^2)\right]x^{i+n-2}+
	\sum_{i} a_{i}x^{i+n}
	=0\\	
& \sum_{i} a_{i}\left( i^2+2in \right)x^{i+n-2}+
	\sum_{i} a_{i}x^{i+n}
	=0\\
\end{split}
\end{equation*}

The equation can be balanced by letting the first sum starts from $i=2$, the second sum from $i=0$ and by setting the coefficient $a_{1}=0$. The equation then becomes:
\begin{equation*}
\sum_{i=2}^{\infty} a_{i}\left( i^2+2in \right)x^{i+n-2}+
	\sum_{i=0}^{\infty} a_{i}x^{i+n}
	=0
\end{equation*}

Now that the equation is balanced, the index shift can be done in order to get the same power of $x$ for the two sums. In particular $k=i-2$ and $k=i$ are set respectively for the first and second sum:
\begin{equation*}
\sum_{k=0}^{\infty} a_{k+2}\left( (k+2)^2+2n(k+2) \right)x^{k+n}+
	\sum_{k=0}^{\infty} a_{k}x^{k+n}
	=0
\end{equation*}

From this equation the recursion formula for the coefficients of the infinite series solution can be defined:
\begin{equation*}
\begin{split}
& a_{k+2}\left( (k+2)^2+2n(k+2) \right)+ a_{k}=0\\
& a_{k+2}\left( (k+2)(2n+k+2) \right)+ a_{k}=0\\
& \quad\to\quad a_{k+2}=-\dfrac{1}{\left( (k+2)(2n+k+2) \right)}a_{k}
\end{split}
\end{equation*}
Since $a_{0}\neq0$, it is possible to define the other coefficients in term of $a_{0}$:
\begin{itemize}
\item $ k=0 $:  $ a_{2}= - \dfrac{1}{2^2\left( 1+n \right)}a_{0} $
\item $ k=2 $:  $ a_{4}= \dfrac{1}{2^4 \cdot 2 \left( 1+n \right)\left( 2+n \right) } a_{0} $
\item ...
\end{itemize}
Therefore, generalizing for $k=2j$ the $2j$-th coefficient becomes:
\begin{equation}
a_{2j}=\dfrac{\left( -1\right)^j}{2^{2j}\cdot j!\left( 1+n \right)\left( 2+n \right)...\left( j+n \right) } a_{0} 
\label{rec3}
\end{equation}

At this point it is convenient to define $a_{0}$ as follow:
\begin{equation*}
a_{0}= \dfrac{1}{2^n \Gamma\left( n+1 \right) }
\end{equation*}
Where $ \Gamma\left(s\right) $ is the $ \Gamma $ function, which is a generalization of the factorial function. Indeed, for $ s $ positive integers:
\begin{equation*}
\Gamma(s+1)=s!
\end{equation*}
Substituting $ a_{0} $ and using the following $ \Gamma $ function property:
\begin{equation*}
\Gamma(x+1)=x\Gamma(x)
\end{equation*}
the denominator of the recursion formula (equation \ref{rec3}) becomes:
\begin{equation}
\begin{split}
2^{2j+n}\cdot j!\Gamma(n+1)\left( 1+n \right)\left( 2+n \right)...\left( j+n \right)&=\\
2^{2j+n}\cdot j!\Gamma(n+2)\left( 2+n \right)...\left( j+n \right) &=\\
2^{2j+n}\cdot j!\Gamma(n+j)\left( j+n \right) &= 2^{2j+n}\cdot j!\Gamma(n+j+1)
\end{split}
\end{equation}

Then, the coefficients can be defined as:
\begin{equation}
a_{2j}=\dfrac{\left( -1\right)^j}{2^{2j+n } j! \Gamma\left( j+1+n \right) } 
\end{equation}
Therefore, the first solution of the radial equation is:
\begin{equation}
R_{1,n}\left( x \right) = \sum_{j=0}^{\infty}\dfrac{\left( -1\right)^j }{ j! \Gamma\left( j+1+n \right) }\left( \dfrac{x}{2}\right)^{2j+n} = J_{n}\left( x\right) 
\label{gensolut}
\end{equation}
where $J_{n}\left( x \right)$ is the $n$-th \textit{Bessel function}.

Looking for the second solution by plugging the generalized infinite series solution, with $r_{2}=-n$, the recurrent relation will be:
\begin{equation}
a_{k+2}=-\dfrac{1}{\left( (k+2)(k+2-2n) \right)}a_{k}
\label{recursion2}
\end{equation}
and the second solution will be:
\begin{equation}
R_{2,-n}\left( x\right)  = \sum_{j=n}^{\infty}\dfrac{\left( -1\right)^j }{ j! \cdot \Gamma\left( j+1-n \right) }\left( \dfrac{x}{2}\right)^{2j-n} = J_{-n}\left( x \right)
\end{equation}
For $n$ integers, this result is not acceptable because the two solutions, $ J_{n}\left( x \right) $ and $ J_{-n}\left( x \right) $, are not linearly independent. Indeed, for $ j\leq n-1 $ the $ \Gamma $ function become negative and it is not defined any more, consequently also the coefficients will be not defined.\\
Therefore, we need to look for a second solution with form: $R_{2}(x)=CR_{1}(x)ln(x)+\sum_{i=0}^{\infty}b_{i}x^{i+r_{2}}$. This second solution is called \textit{Bessel function of the second kind}, or \textit{Neumann function} (in the case of $n$ integer), $N_{n}(x)$:
\begin{equation}
N_{n}( x ) = \dfrac{J_{n}(x)cos(\pi n)-J_{-n}(x)}{sin(\pi n)}
\end{equation}
Therefore, recalling that $ x=\beta\rho $, the complete solution of the radial equation will be:
\begin{equation}
R_{n}\left( \beta \rho \right) = A_{n}J_{n}\left(\beta\rho\right) +B_{n}N_{n}\left( \beta \rho \right)=A_{n}J_{n}\left(\beta\rho\right)
\end{equation}
\begin{figure}[t]
\centering
\makebox[\textwidth]{
\subfloat[]
{\includegraphics[width=6cm]{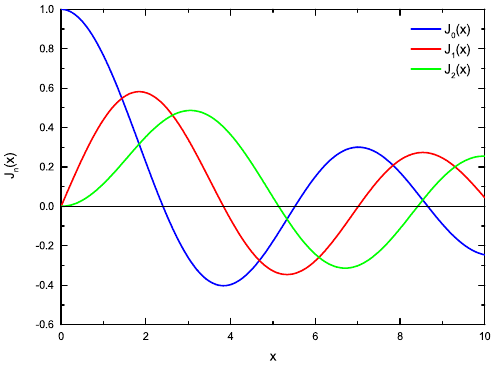}\label{Jn}}
\hspace{3mm}
\subfloat[]
{\includegraphics[width=6cm]{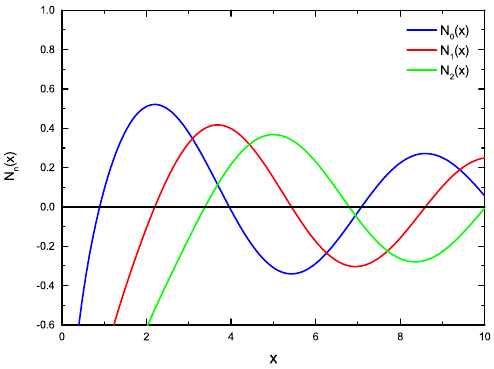}\label{Nn}}
}
\caption{a) First three orders Bessel functions. b) First three orders Naumann functions.}
\label{JnNn}
\end{figure}

Since the Neumann function is not finite at $r=0$, it is not a physically acceptable solution inside a cylindrical conductor (this is an internal problem), and we set $B_{n}=0$. This different behaviour of Bessel and Neumann functions is shown in figure \ref{JnNn}.\\
So, the complete general solution of the wave equation, before imposing boundary conditions, is:
\begin{equation}
u(\rho,\varphi,z,t)=A_{n}J_{n}(\beta\rho)\left[ Asin(\kappa z)+Bcos(\kappa z)\right]e^{\pm in \varphi} e^{\pm i \omega t}
\label{usolution}
\end{equation}
with $ n=0,1,2... $.

Now we need to figure out the boundary conditions for the general wave equation solution for both $E_{z}$ and $H_{z}$. Considering the interface between the perfect conductor and the vacuum, the tangential component of the electric field $\mathbf{E}$, and the normal component of the magnetic field $\mathbf{H}$ must be continuous across the boundary. Then, the following boundary conditions at the cylindrical surface are required:
\begin{equation}
\begin{split}
\hat{n} \times \mathbf{E}=0, \quad \hat{n} \cdot \mathbf{H}=0
\end{split}
\end{equation}
where $ \hat{n} $ is the unit vector normal to the surface $ S $. Then the boundary conditions for the fields components along the $ z $-axis are:
\begin{equation}
\begin{split}
E_{z}\vert_{S}=0, \quad \dfrac{\partial H_{z}}{\partial \hat{n}}\bigg|_{S}=0
\end{split}
\label{boundCly}
\end{equation}
where $\partial / \partial \hat{n}$ is the normal derivative at a point on the curved surface $S$ of the cylinder. Since we have different boundary conditions for $E_{z}$ and $H_{z}$, there will be two type of field configuration inside the cylinder: $TM$ (transverse magnetic) and $TE$ (transverse electric), as expected from section \ref{S:2}. So, the general solution of the wave equation (equation \ref{usolution}) has to be adopted to both of these two configuration separately.

\subsection{Transverse Magnetic Modes in a Cylindrical Cavity}
\label{Ss:1}
\begin{figure}[t]
\begin{center}
\includegraphics[scale=0.4]{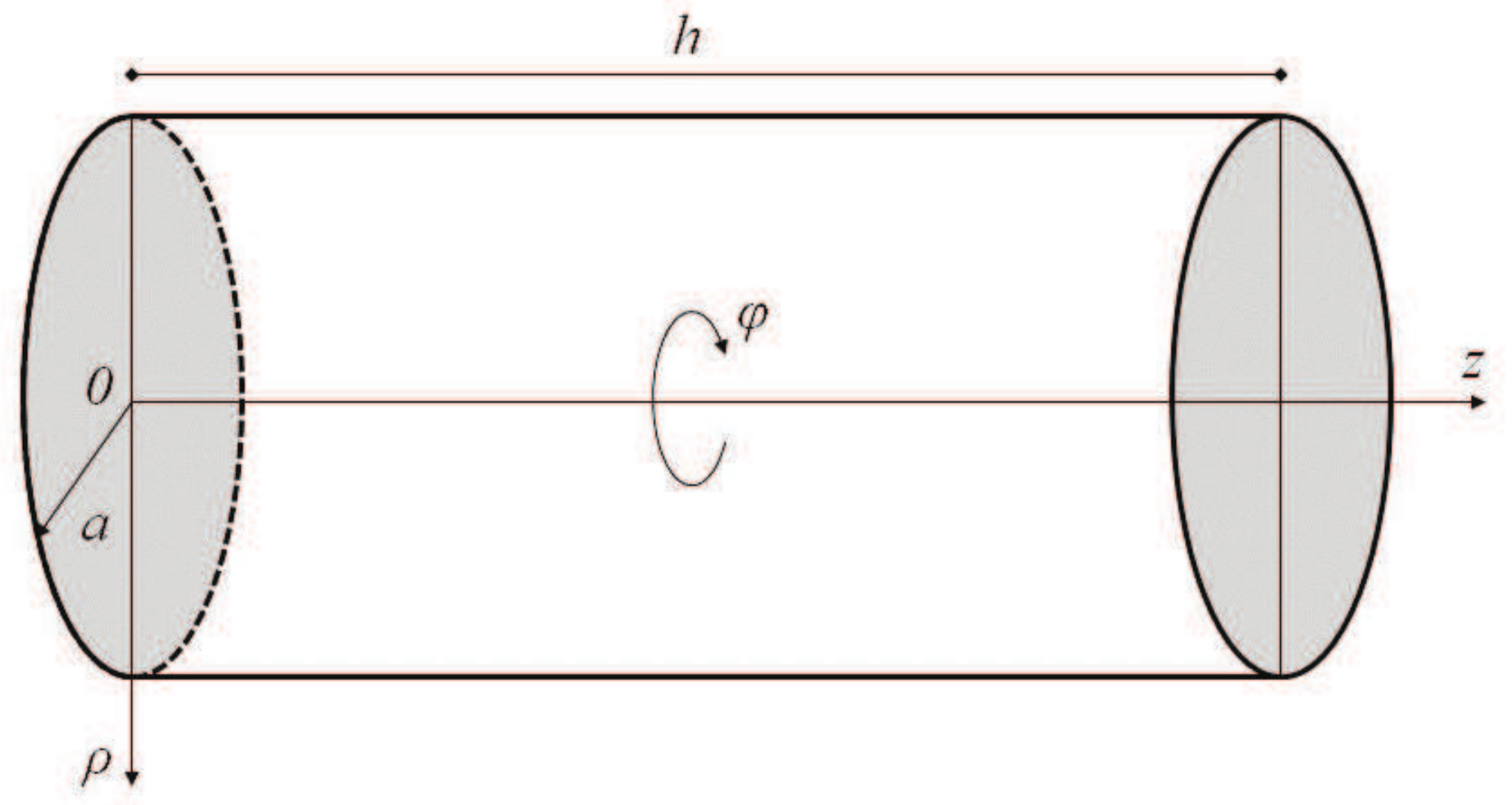}
\caption{Sketch of a cylindrical ``pill box'' cavity.}
\label{fig:Cav}
\end{center}
\end{figure}

The cylindrical cavity resonator, pill-box (figure \ref{fig:Cav}), is a hollow cylindrical wave-guide of radius $a$ with two end caps at $ z=0 $ and $ z=h $. It is assumed that these conducting surfaces are planar and perpendicular to the axis of the cylinder.

The boundary conditions for the $TM$ modes are:
\begin{equation}
H_{z}=0 \quad $everywhere,$ \quad E_{z}\vert_{\rho=a}=0$,$\quad E_{t}\vert_{z=0,h}=0
\end{equation}

This means that the electric field along the $z$-axis and the tangential one should vanish at the curved surface and at the two end caps respectively. Then, replacing $ u $ with $ E_{z} $ and $ A_{n} $ with $ E_{0} $ in equation \ref{usolution}, and assuming a fix radius $\rho=a$ we have:
\begin{equation}
E_{z}(a,\varphi,z,t)=E_{0}J_{n}(\beta a)\left[ Asin(\kappa z)+Bcos(\kappa z)\right]e^{\pm in \varphi} e^{\pm i \omega t}=0
\end{equation}
hence in order to respect the boundary conditions we have:
\begin{equation}
J_{n}(\beta a)=0 \quad \to \quad \beta a=\alpha_{n,l} \quad \to \quad \beta=\dfrac{\alpha_{n,l}}{a} 
\label{zeroB}
\end{equation}
where $ l=1,2,3,... $ and $ \alpha_{nl} $ identifies the $ l $-th  zero of the $ n $-th  Bessel function $ J_{n} $.\\

As defined, the tangential component of the electric field should vanish at the end caps, then: $ E_{t}=0  $ at the end caps, so from the Maxwell equation $ \nabla \cdot E=0 $ follows that:
\begin{equation}
\begin{split}
\nabla_{t}E_{t}+\dfrac{\partial E_{z}}{\partial z}=0 \quad \to \quad \dfrac{\partial E_{z}}{\partial z}=0
\end{split}
\end{equation}
Then, the solution should satisfy the boundary conditions:
\begin{equation}
\dfrac{\partial E_{z}(\rho,\varphi,0,t)}{\partial z}=\dfrac{\partial E_{z}(\rho,\varphi,h,t)}{\partial z}=0
\end{equation}
so:
\begin{equation}
\begin{split}
\dfrac{\partial E_{z}(\rho,\varphi,0,t)}{\partial z}&=E_{0}J_{n}\left(\dfrac{\alpha_{nl}}{a} \rho\right)Ake^{\pm in \varphi} e^{\pm i \omega t}=0\\
\quad &\to A=0 \\ \quad
\dfrac{\partial E_{z}(\rho,\varphi,h,t)}{\partial z}&=E_{0}J_{n}\left(\dfrac{\alpha_{nl}}{a} \rho\right)\left[ -Bksin(\kappa h)\right]e^{\pm in \varphi} e^{\pm i \omega t}=0\\
\quad &\to \kappa_{m} = \dfrac{m\pi}{h} \quad with \quad m=0,1,2,...
\end{split}
\label{quantk}
\end{equation}

Therefore, the electric field along the $z$-axis for $TM$ modes in a pill-box cavity will be:
\begin{equation}
E_{z}(\rho,\varphi,z,t)=E_{0}J_{n}\left(\dfrac{\alpha_{nl}}{a} \rho\right)cos\left(\dfrac{m\pi}{h} z\right)e^{\pm in \varphi} e^{\pm i \omega_{nlm} t}
\label{gensolpb}
\end{equation}
with $ n=0,1,2,... $, $ l=1,2,3,... $ and $ m=0,1,2,... $.\\
And the eigenvalues are:
\begin{equation}
\beta=\sqrt{q^2-\kappa^2}=\sqrt{\dfrac{\omega^2}{c^2}-\dfrac{m^2\pi^2}{h^2}}
\label{TMeigenvalue}
\end{equation}

The general solution we found for the angular part (equation \ref{phisolution}) is a linear combination of sine and cosine in terms of $ \varphi $, but now it becomes more convenient to rewrite it only in terms of cosine; then, the solution for $ E_{z} $ is:
\begin{equation}
E_{z}(\rho,\varphi,z,t)=E_{0}J_{n}\left(\dfrac{\alpha_{nl}}{a} \rho\right)cos\left(\dfrac{m\pi}{h} z\right)cos(n \varphi) e^{\pm i \omega_{nlm} t}
\label{solEzTM}
\end{equation}
In addition, combining equations \ref{zeroB} and \ref{TMeigenvalue} we found:
\begin{equation}
\dfrac{\alpha_{nl}}{a}=\sqrt{\dfrac{\omega^2}{c^2}-\dfrac{m^2\pi^2}{h^2}}
\end{equation}
which can be rewritten in terms of the resonant frequency:
\begin{equation}
\omega_{nlm}=c\sqrt{\left(\dfrac{\alpha_{nl}}{a}\right)^2+\left(\dfrac{m\pi}{h}\right)^2}
\label{EqfreqTM}
\end{equation}
with $ n=0,1,2,... $, $ l=1,2,3,... $ and $ m=0,1,2,... $.

\begin{figure}[t]
\centering
\makebox[\textwidth]{
\subfloat[]
{\includegraphics[width=4.6cm]{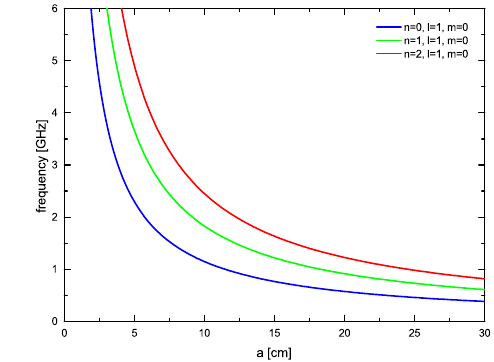}\label{Freq1TM}}
\hspace{2mm}
\subfloat[]
{\includegraphics[width=4.6cm]{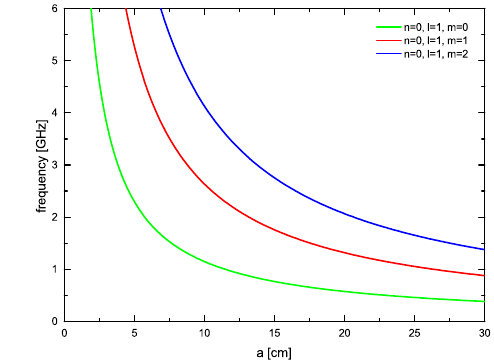}\label{Freq2TM}}
\hspace{2mm}
\subfloat[]
{\includegraphics[width=4.6cm]{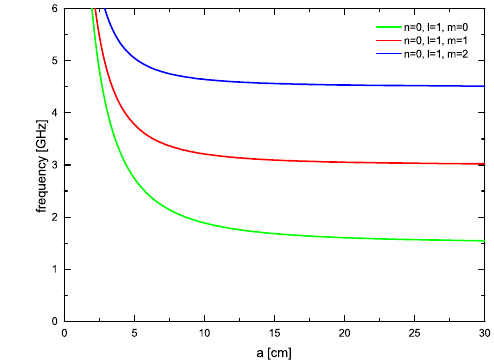}\label{Freq3TM}}
}
\caption{a) Resonant frequency as a function of the cavity radius for different values of $n$, with $l=1$ and $m=0$. b) Resonant frequency as a function of the cavity radius for different values of $l$, with $n=0$ and $m=0$. c) Resonant frequency as a function of the cavity radius for different values of $m$, with $n=0$ and $l=1$.}
\label{FreqTM}
\end{figure}

In figure \ref{FreqTM} the resonance frequency as a function of cavity radius is plotted, setting the cavity length at $ h=10cm $. It can be seen that the resonance frequency decrease with the cavity radius, this reduction is prominently for small values of radius $a$, and it becomes smoother as the radius increases tending to an asymptote defined by the index $m$. Indeed, $\lim_{x\to\infty}\omega_{nlm}= m\pi /h$, as it can be deduced from equation \ref{EqfreqTM}.

In figure \ref{Freq1TM} the curves are plotted for different values of $n$, with $l=1$ and $m=0$.  In figure \ref{Freq2TM} the curves are plotted for different values of $l$, with $n=0$ and $m=0$. In figure \ref{Freq3TM} the curves are plotted for different values of $m$, with $n=0$ and $l=1$. The lowest resonant frequency values are given, in all the three cases by the smallest indexes: $n=0$ for fig. \ref{Freq1TM}, $ l=1 $ for fig. \ref{Freq2TM}, and $ m=0 $ for fig. \ref{Freq3TM}. It can be also seen that the resonant frequency increases as the three indexes increase, but, increasing $ l $ (\ref{Freq2TM}) the effect is more pronounced for low values of cavity radius, whereas increasing $ m $ (\ref{Freq3TM}) this effect is more pronounced for high values of cavity radius.

The other components of the electric field and magnetic field can be calculated by using equations \ref{TM}:
\begin{equation}
\begin{split}
	\mathbf{E_{t}}&=\pm\dfrac{i\kappa}{\beta^2}\nabla_{t}E_{z}\\
	\mathbf{H_{t}}&=\dfrac{i\omega\mu_{0}}{\beta^2}\hat{z}\times\nabla_{t}E_{z}
\end{split}
\end{equation}
these equations were found out by substituting the derivative of the fields respect $z$ with $\pm i\kappa$, since the fields was chosen to have a $ z $ dependence of the type $ e^{\pm i\kappa z} $. Now the $z$ field dependence is of the type $ cos(\kappa z) $, then:
\begin{equation}
\dfrac{\partial E_{z}}{\partial z}=-\dfrac{m\pi}{h}sin\left(\dfrac{m\pi}{h}z\right)Y(\rho,\varphi)
\end{equation}
where the function $ Y(\rho,\varphi) $ contains the dependence of $ \rho $ and $ \varphi $ of the field $ E_{z} $ (see eq. \ref{eq:YZT}), indeed:
\begin{equation}
E_{z}=Y(\rho, \varphi)cos(\kappa_{m z}) $ , $\quad
Y(\rho, \varphi)= E_{0}J_{n}\left(\dfrac{\alpha_{nl}}{a} \rho\right)cos(n\varphi)
\label{angular}
\end{equation}

Therefore, equations \ref{TM} becomes:
\begin{equation}
\begin{split}
\mathbf{E_{t}}&=-\dfrac{m\pi}{h\beta^2}sin\left(\dfrac{m\pi z}{h} \right)\nabla_{t}\mathbf{Y}(\rho,\varphi)\\
\quad &=-\dfrac{m\pi}{h\beta^2}sin\left(\dfrac{m\pi z}{h} \right)\left[ \hat{\rho}\dfrac{\partial}{\partial\rho} + \dfrac{\hat{\varphi}}{\rho}\dfrac{\partial}{\partial\varphi}\right]Y(\rho,\varphi) \\ \quad
\mathbf{H_{t}}&=\dfrac{i\varepsilon_{0} \omega}{\beta^2}cos\left(\dfrac{m\pi z}{h} \right)\hat{z} \times \nabla_{t} \mathbf{Y}(\rho,\varphi)\\
\quad &=\dfrac{i\varepsilon_{0} \omega}{\beta^2}cos\left(\dfrac{m\pi z}{h} \right)\hat{z} \times \left[ \hat{\rho}\dfrac{\partial}{\partial\rho} + \dfrac{\hat{\varphi}}{\rho}\dfrac{\partial}{\partial\varphi}\right]Y(\rho,\varphi) \\
\end{split}
\end{equation}

Now, recalling that the derivative of the $ n $-th Bessel function is:
\begin{equation}
J^{'}_{n}(x)= \left(\dfrac{n}{x}\right)J_{n}
\left(x\right)-J_{n+1}\left(x\right)
\label{BessDerivx}
\end{equation}
replacing $ x=\beta\rho $ we obtain:
\begin{equation}
J^{'}_{n}(\beta \rho)=\beta \left[
\left(\dfrac{n}{\rho \beta}\right)
J_{n}\left(\beta \rho\right)-J_{n+1}\left(\beta \rho\right)
\right]
\label{BessDeriv}
\end{equation}
then, the derivatives of $ Y(\rho,\varphi) $ give the following results:
\begin{equation}
\begin{split}
&\dfrac{\partial Y(\rho,\varphi)}{\partial\rho}=\beta E_{0}
\left[
\left(\dfrac{na}{\rho\alpha_{nl}}\right)
J_{n}\left(\dfrac{\alpha_{nl}}{a} \rho\right)- J_{n+1}\left(\dfrac{\alpha_{nl}}{a} \rho\right)
\right]
cos(n\varphi)\\ 
\quad &\dfrac{\partial Y(\rho,\varphi)}{\partial\varphi}=-nE_{0}J_{n}\left(\dfrac{\alpha_{nl}}{a} \rho\right)sin(n\varphi)
\end{split}
\end{equation}

And we found that the complete set of $ TM_{nlm} $ modes are:
\begin{equation}
\begin{split}
E_{z}&=E_{0}J_{n}\left( \dfrac{\alpha_{nl}}{a} \rho\right) cos\left( \dfrac{m\pi z}{h}\right) cos\left( n \varphi\right) e^{\pm i \omega_{nlm} t}\\
E_{\rho}&=-\dfrac{m\pi a}{h \alpha_{nl}}E_{0}
\left[
\left(\dfrac{na}{\rho\alpha_{nl}}\right)
J_{n}\left(\dfrac{\alpha_{nl}}{a} \rho\right)- J_{n+1}\left(\dfrac{\alpha_{nl}}{a} \rho\right)
\right]\\
&\quad\quad\quad\quad\quad\quad\cdot sin\left( \dfrac{m\pi z}{h}\right)cos\left( n \varphi\right)  e^{\pm i \omega_{nlm} t}\\
\quad 
E_{\varphi}&=\dfrac{m n\pi a^2}{\rho h \alpha_{nl}^{2}}E_{0}J_{n}\left(\dfrac{\alpha_{nl}}{a}  \rho\right) cos\left( \dfrac{m\pi z}{h}\right) sin\left( n \varphi\right)  e^{\pm i \omega_{nlm} t}\\
\quad 
H_{z}&=0\\
\quad 
H_{\rho}&=\dfrac{i\varepsilon_{0}\omega_{nlm} n a^2}{\rho \alpha_{nl}^{2}}E_{0}J_{n}\left(\dfrac{\alpha_{nl}}{a} \rho\right) cos\left( \dfrac{m\pi z}{h}\right) sin\left( n \varphi\right)  e^{\pm i \omega_{nlm} t}\\
\quad 
H_{\varphi}&=\dfrac{i\varepsilon_{0}\omega_{nlm} a}{\alpha_{nl}}E_{0}
\left[
\left(\dfrac{na}{\rho\alpha_{nl}}\right)
J_{n}\left(\dfrac{\alpha_{nl}}{a} \rho\right)- J_{n+1}\left(\dfrac{\alpha_{nl}}{a} \rho\right)
\right]\\
&\quad\quad\quad\quad\quad\quad\cdot cos\left( \dfrac{m\pi z}{h}\right)cos\left( n \varphi\right)  e^{\pm i \omega_{nlm} t}\\
\end{split}
\end{equation}
these modes are usually classified by the nomenclature $ TM_{nlm} $, where the integers $ n $, $ l $ and $ m $ measures the number of the sign changes of the field along the $ \varphi $, $ \rho $, and $ z $ directions respectively. Table \ref{ZeroBTab1} shows some zeros $ \alpha_{nl} $ of the $n$-th Bessel function.
\begin{table}
\centering
\caption{Zeros of the $n$-th Bessel function.}
\begin{tabular}[c]{ c c c c c }
\hline
$l$ & $J_{0}(x)$ & $J_{1}(x)$ & $J_{2}(x)$ & $J_{3}(x)$ \\
\hline
1   &   2.4048   &   3.8317   &   5.1356   &   6.3802 \\
2   &   5.5201   &   7.0156   &   8.4172   &   9.7610 \\
3   &   8.6537   &   10.1735  &   11.6198  &   13.0152\\ 
4   &   11.7915  &   13.3237  &   14.7960  &   16.2235 \\
\hline
\end{tabular} 
\label{ZeroBTab1}		
\end{table}

The lowest $TM$ mode (fundamental mode) has $n=0$, $l=1$ and $m=0 $, then it is denoted as $TM_{010}$. The resonance frequency is given by:
\begin{equation}
\omega_{010}=\dfrac{2.405c}{a}
\end{equation}
and the fields are:
\begin{equation}
\begin{split}
E_{z}&=E_{0}J_{0}\left( \dfrac{2.405}{a} \rho\right)e^{\pm i \omega_{010} t}\\ 
\quad
H_{\varphi}&=-i\varepsilon_{0} c E_{0}J_{1}\left( \dfrac{2.405}{a} \rho\right)e^{\pm i \omega_{010} t}\\ 
\end{split}
\end{equation}

It is possible to notice that the resonance frequency of the mode $TM_{010}$ is independent from the cavity length, and it depends only on its radius $a$. This results is not surprising since the fields vary along  the $\rho$ direction but not along $z$.

\begin{figure}[p]
\centering
\makebox[\textwidth]{
\subfloat[$TM_{010}$]
{\includegraphics[scale=0.75]{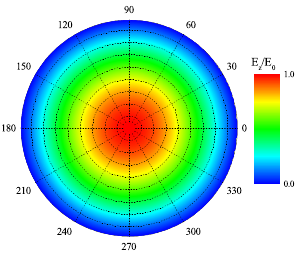}\label{TM010}}
\hspace{2mm}
\subfloat[$TM_{020}$]
{\includegraphics[scale=0.75]{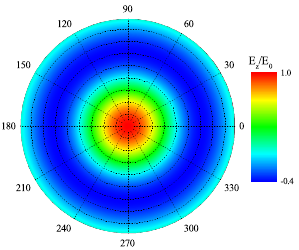}\label{TM020}}
\hspace{2mm}
\subfloat[$TM_{030}$]
{\includegraphics[scale=0.75]{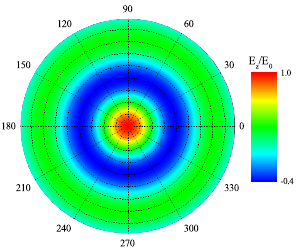}\label{TM030}}
}
\vspace{3mm}
\centering
\makebox[\textwidth]{
\subfloat[$TM_{110}$]
{\includegraphics[scale=0.75]{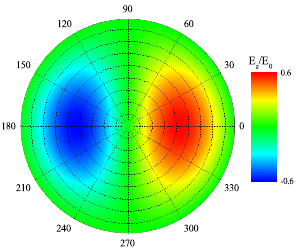}\label{TM110}}
\hspace{2mm}
\subfloat[$TM_{120}$]
{\includegraphics[scale=0.75]{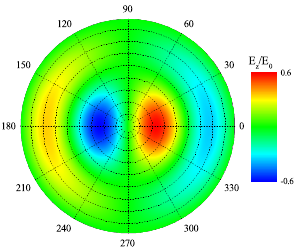}\label{TM120}}
\hspace{2mm}
\subfloat[$TM_{130}$]
{\includegraphics[scale=0.75]{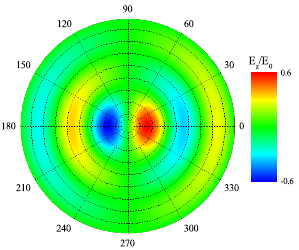}\label{TM130}}
}
\vspace{3mm}
\centering
\makebox[\textwidth]{
\subfloat[$TM_{210}$]
{\includegraphics[scale=0.75]{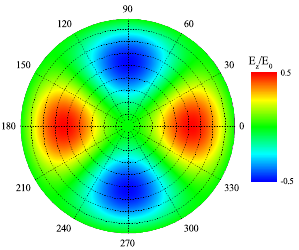}\label{TM210}}
\hspace{2mm}
\subfloat[$TM_{220}$]
{\includegraphics[scale=0.75]{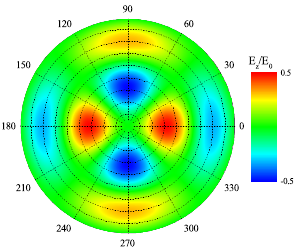}\label{TM220}}
\hspace{2mm}
\subfloat[$TM_{230}$]
{\includegraphics[scale=0.75]{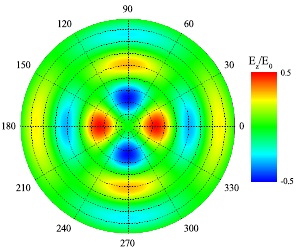}\label{TM230}}
}
\vspace{3mm}
\centering
\makebox[\textwidth]{
\subfloat[$TM_{310}$]
{\includegraphics[scale=0.75]{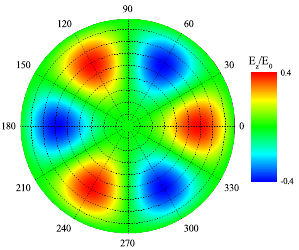}\label{TM310}}
\hspace{2mm}
\subfloat[$TM_{320}$]
{\includegraphics[scale=0.75]{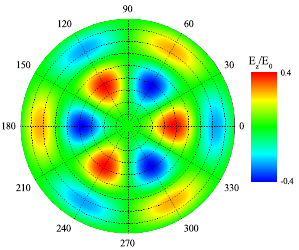}\label{TM320}}
\hspace{2mm}
\subfloat[$TM_{330}$]
{\includegraphics[scale=0.75]{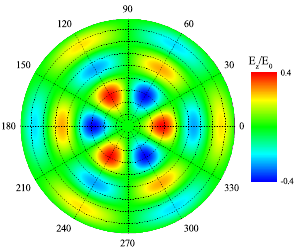}\label{TM330}}
}
\caption{Polar plots of $E_{z}$ for $m=0$ and different couples of $n$ and $l$ at $z=h/2$.}
\label{polarTM}
\vspace*{\floatsep}
\centering
\makebox[\textwidth]{
\subfloat[$TM_{010}$]
{\includegraphics[scale=0.75]{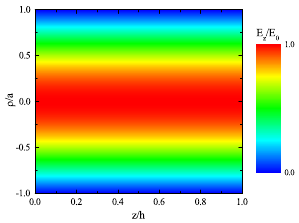}\label{TM010b}}
\hspace{2mm}
\subfloat[$TM_{020}$]
{\includegraphics[scale=0.75]{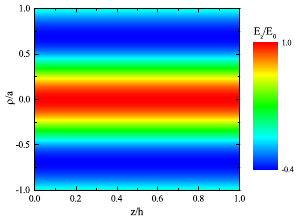}\label{TM020b}}
\hspace{2mm}
\subfloat[$TM_{110}$]
{\includegraphics[scale=0.75]{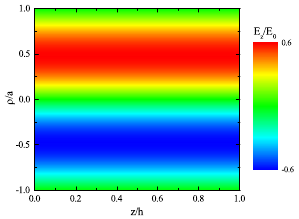}\label{TM110b}}
}
\vspace{3mm}
\centering
\makebox[\textwidth]{
\subfloat[$TM_{011}$]
{\includegraphics[scale=0.75]{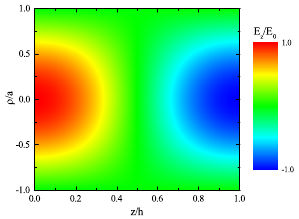}\label{TM011}}
\hspace{2mm}
\subfloat[$TM_{021}$]
{\includegraphics[scale=0.75]{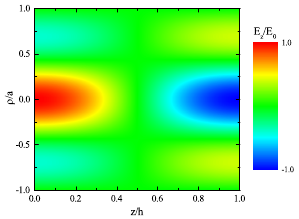}\label{TM021}}
\hspace{2mm}
\subfloat[$TM_{111}$]
{\includegraphics[scale=0.75]{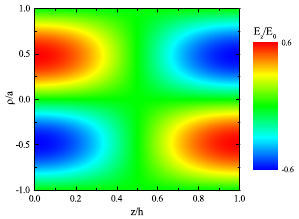}\label{TM111}}
}
\vspace{3mm}
\centering
\makebox[\textwidth]{
\subfloat[$TM_{012}$]
{\includegraphics[scale=0.75]{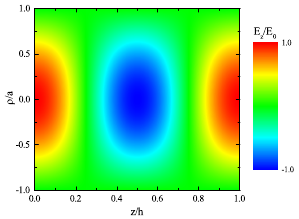}\label{TM012}}
\hspace{2mm}
\subfloat[$TM_{022}$]
{\includegraphics[scale=0.75]{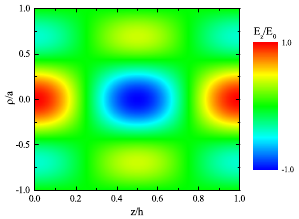}\label{TM022}}
\hspace{2mm}
\subfloat[$TM_{112}$]
{\includegraphics[scale=0.75]{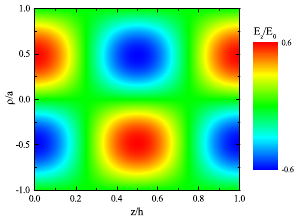}\label{TM112}}
}
\vspace{3mm}
\centering
\makebox[\textwidth]{
\subfloat[$TM_{013}$]
{\includegraphics[scale=0.75]{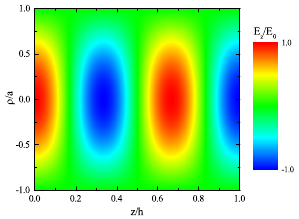}\label{TM013}}
\hspace{2mm}
\subfloat[$TM_{023}$]
{\includegraphics[scale=0.75]{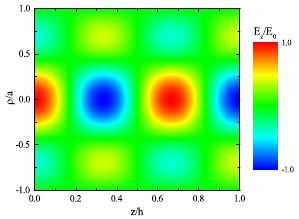}\label{TM023}}
\hspace{2mm}
\subfloat[$TM_{113}$]
{\includegraphics[scale=0.75]{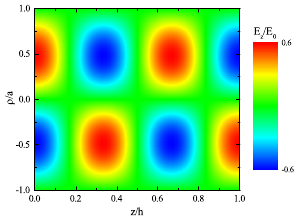}\label{TM113}}
}
\caption{Axial plots of $E_{z}$ for different combinations of $n$, $l$ and $m$ at $\varphi=0$.}
\label{axialTM}
\end{figure}

The trend of the fields can be clearly seen from figures \ref{TM010} and \ref{TM010b}, in which the $z$ component of the electric field is plotted for the $TM_{010}$ mode.

In figure \ref{TM010} the polar plot of $E_{z}/E_{0}$ is shown for $z=h/2$. It can be noticed that the intensity of the field is the highest at the center of the cavity ($\rho=0$), and decreases by approaching the cavity walls, where $\rho=a$. While, from the axial plot in figure \ref{TM010b}, it is clear that the field is constant along the cavity length ($z$-axis). Therefore, as expected, for $n=0$, $l=1$ and $m=0$ there are no sign changes of the fields along any directions.

In figure \ref{polarTM} are also shown the polar plots of other $TM_{nl0}$ modes, all calculated at the middle of the cavity ($z=h/2$).

Comparing the modes $TM_{010}$, $TM_{020}$ and $TM_{030}$ (figures \ref{TM010}, \ref{TM020}, \ref{TM030}) we can easily identify how the quantum index $l$ quantizes the field along the radius of the cavity. As expected, when $l=1$ the field does not change sign along the $\rho$ direction, whereas the field changes sign one time when $l=2$, and two times when $l=3$.

The quantization introduced by the index $n$ can be observed by comparing the modes $TM_{010}$, $TM_{110}$, $TM_{210}$ and $TM_{310}$ plotted in figures \ref{TM010}, \ref{TM110}, \ref{TM210} and \ref{TM310}. When $n=0$ the field does not change sign along the $\varphi$ direction, whereas with $n=1$ the field changes sign one time, with $n=2$ two times, and with $n=3$ three times.

Then, in order to highlight how the index $m$ acts, we can for example comparing the modes $TM_{010}$, $TM_{011}$, $TM_{012}$ and $TM_{013}$ plotted in figures \ref{TM010b}, \ref{TM011}, \ref{TM012} and \ref{TM013}. When $m=0$ the field does not change sign along the $z$ direction, whereas the field changes sign one time when $m=1$, two times when $m=2$ and three times when $m=3$.

It is important to underline that all $TM_{0lm}$ modes are suitable to accelerate particles, indeed they are the only ones that have a non-zero electric field along the $z$ direction for $\rho=0$ (which is also the beam axis). This is intrinsically introduced by the Bessel function. Indeed, recalling figure \ref{Jn} only $J_{0}$ is non zero at $\rho=0$. Anyhow, the mode that is universally used to accelerate particles with $TM$ class cavities (as the ``pill box'') is the $TM_{010}$ the fundamental one, i.e. the one with the lowest frequency and the highest electric field along the beam axis.

\newgeometry{left=2cm,right=2cm,bottom=1.5cm,top=0.5cm}
\pagestyle{empty}
\restoregeometry
\pagestyle{plain}

Another interesting class of modes is the $TM_{1lm}$. This kind of resonant fields are called bipolar because the field changes sign one time around $\varphi$. Along the $z$ direction the electric field is present simultaneously with opposite signs, as it can be noticed from figure \ref{TM110}. This field configuration implies a net value of magnetic field along $\rho$ and $\varphi=90°$ or $180°$. This oscillating magnetic field is often used in accelerator colliders to impart a transversal momentum to the particles bunch to maximize the luminosity at the interaction point of the two particle beams.

\subsection{Transverse Electric Modes in a Cylindrical Cavity}
\label{Ss:2}
The boundary conditions for the $TE$ modes are:
\begin{equation}
E_{z}=0 \quad $ everywhere,$ \quad \dfrac{\partial H_{z}}{\partial \hat{n}}\bigg|_{\rho=a}=0$,$\quad  H_{z}\vert_{z=0,h}=0
\end{equation}

This means that the derivative of the $z$ component of the magnetic field, $H_{z}$, respect the normal to the curved surface, should vanish at $\rho=a$. Also, $H_{z}$ should be equal to zero at the end caps.\\
The function $ Y(\rho,\varphi) $ contains the dependence of $ \rho $ and $ \varphi $ of the field $ H_{z} $, indeed: $ H_{z}=Y(\rho, \varphi)cos(\kappa_{m z}) $ (see eq. \ref{eq:YZT}). The magnetic field along the $z$ direction $ H_{z}$ is defined by replacing $ u $ with $ H_{z}$ and $ A_{n} $ with $ H_{0} $ in equation \ref{usolution}:
\begin{equation}
H(\rho,\varphi,z)=H_{0}J_{n}(\beta\rho)\left[ Asin(\kappa z)+Bcos(\kappa z)\right]e^{\pm in \varphi}
\label{HsolutionTE}
\end{equation}
Therefore, $Y(\rho,\varphi)$ can be defined as:
\begin{equation}
Y(\rho, \varphi)= H_{0}J_{n}(\beta \rho)e^{\pm in \varphi}
\label{angular2}
\end{equation}
Where $A_{n}$ is set equal to $H_{0}$.

Therefore, applying the boundary conditions:
\begin{equation}
\dfrac{\partial Y(a,\varphi,z,t)}{\partial \rho}=H_{0}J^{'}_{n}(\beta a)e^{\pm in \varphi}=0
\end{equation}
it should be:
\begin{equation}
J^{'}_{n}(\beta a)=0 \quad \to \quad \beta a=\alpha^{'}_{n,l} \quad \to \quad \beta=\dfrac{\alpha^{'}_{n,l}}{a} \quad with \quad l=1,2,3,...
\label{zeroB2}
\end{equation}
where $ \alpha^{'}_{nl} $ identifies the $ l $-th  zero of the $ n $-th  $ J^{'}_{n} $ Bessel function.

In addition, the magnetic field along the $z$-axis should vanish at the end caps, then: $ H_{z}(\rho,\varphi,0,t)=H_{z}(\rho,\varphi,h,t)=0  $. So:
\begin{equation}
\begin{split}
H_{z}(\rho,\varphi,0,t)&=H_{0}J_{n}\left(\dfrac{\alpha^{'}_{nl}}{a}\rho\right)Be^{\pm in \varphi} e^{\pm i \omega_{nlm} t}=0 \\ \quad 
&\to \quad B=0\\
\quad H_{0}(\rho,\varphi,h,t)&=H_{0}J_{n}\left(\dfrac{\alpha^{'}_{nl}}{a}\rho\right)\left[Asin(\kappa h)\right]e^{\pm in \varphi} e^{\pm i \omega_{nlm} t}=0\\ \quad 
&\to \quad \kappa_{m}=\dfrac{m\pi}{h} \quad with \quad m=0,1,2,...
\end{split}
\label{quantk2}
\end{equation}

Therefore, the magnetic field along the $z$-axis for a $TE$ mode in a pill-box cavity will be:
\begin{equation}
H_{z}(\rho,\varphi,z,t)=H_{0}J_{n}\left(\dfrac{\alpha^{'}_{nl}}{a} \rho\right)sin\left(\dfrac{m\pi}{h} z\right)e^{\pm in \varphi} e^{\pm i \omega_{nlm} t}
\end{equation}
with $ n=0,1,2,... $, $ l=1,2,3,... $ and $ m=1,2,3,... $, because $ m=0 $ is the trivial solution.\\
Also in this case, we introduce a more convenient dependence of $ \varphi $ in terms of cosine:
\begin{equation}
H_{z}(\rho,\varphi,z,t)=H_{0}J_{n}\left(\dfrac{\alpha^{'}_{nl}}{a} \rho\right)sin\left(\dfrac{m\pi}{h} z\right)cos\left(n \varphi\right) e^{\pm i \omega_{nlm} t}
\end{equation}
In addition, combining equations \ref{zeroB2} and \ref{quantk2}, we found the eigenvalues $ \beta $:
\begin{equation}
\beta=\dfrac{\alpha^{'}_{nl}}{a}=\sqrt{q^2-\kappa^2}=\sqrt{\dfrac{\omega^2}{c^2}-\dfrac{m^2\pi^2}{h^2}}
\end{equation}
this equation can be rewritten in terms of the resonant frequency for the $TE$ modes:
\begin{equation}
\omega_{nlm}=c\sqrt{\left(\dfrac{\alpha^{'}_{nl}}{a}\right)^2+\left(\dfrac{m\pi}{h}\right)^2}
\label{Eqfreq2}
\end{equation}
with $ n=0,1,2,... $, $ l=1,2,3,... $ and $ m=1,2,3,... $.

\begin{figure}[t]
\centering
\makebox[\textwidth]{
\subfloat[]
{\includegraphics[width=4.6cm]{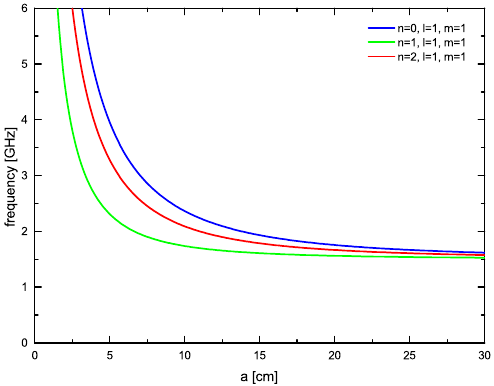}\label{Freq1TE}}
\hspace{2mm}
\subfloat[]
{\includegraphics[width=4.6cm]{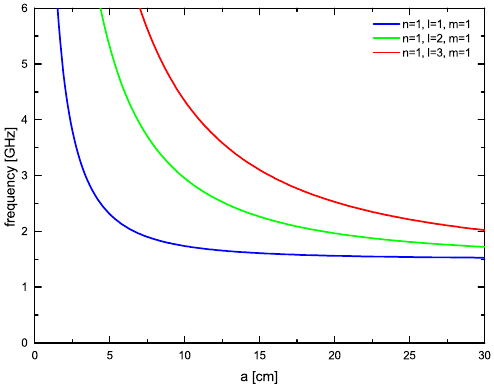}\label{Freq2TE}}
\hspace{2mm}
\subfloat[]
{\includegraphics[width=4.6cm]{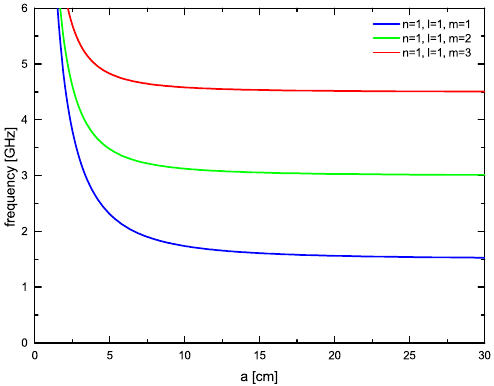}\label{Freq3TE}}
}
\caption{a) Resonant frequency as a function of the cavity radius for different values of $n$, with $l=1$ and $m=1$. b) Resonant frequency as a function of the cavity radius for different values of $l$, with $n=1$ and $m=1$. c) Resonant frequency as a function of the cavity radius for different values of $m$, with $n=1$ and $l=1$.}
\label{FreqTE}
\end{figure}

In figure \ref{FreqTE} is shown the dependence of the resonance frequency on the radius of the cavity for different combination of the three indexes $n$, $l$ and $m$ for the $TE$ modes, for a cavity length equal to $h=10cm$.

In all the three graphs similar trends of the frequency with the radius can be observed. In particular, as the radius increases the frequency decreases, tending to an asymptote defined by the index $m$. Indeed, $\lim_{x\to\infty}\omega_{nlm}= m\pi /h$, as it can be deduced from equation \ref{Eqfreq2}.

However, for different combinations of indexes the shape of the curve changes. In figure \ref{Freq1TE} only the index $n$ changes, while the other two are fixed to $l=1$ and $m=1$. Surprising, it can be seen that the lowest frequency mode is the $TE_{111}$ and not the $TE_{011}$, which has instead a higher frequency. Therefore, the fundamental $TE$ mode inside a cylindrical cavity is the $TE_{111}$.

In figure \ref{Freq2TE} are plotted the curves keeping fixed the indexes $n=1$ and $m=1$, while $l$ is free to change. In this case it can be seen that as the index $l$ increases the slope of the curve decreases, implying higher frequencies for the same radius.

In figure \ref{Freq3TE} is instead highlighted how the minimum asymptotic frequency is dependent on the index $m$, as expected the higher $m$ the higher the minimum frequency is.

The other components of the electric field and magnetic field for the $TE$ modes can be calculated by using the following relations (see equation \ref{TE}):
\begin{equation}
\begin{split}
\mathbf{E_{t}}&=-\dfrac{i\omega\mu_{0}}{\beta^2}sin\left(\dfrac{m\pi z}{h} \right)\hat{z} \times \nabla_{t}\mathbf{Y}(\rho,\varphi)\\
\quad &=-\dfrac{i\omega\mu_{0}}{\beta^2}sin\left(\dfrac{m\pi z}{h} \right)\hat{z} \times\left[ \hat{\rho}\dfrac{\partial}{\partial\rho} + \dfrac{\hat{\varphi}}{\rho}\dfrac{\partial}{\partial\varphi}\right]Y(\rho,\varphi) \\ \quad
\mathbf{H_{t}}&=\dfrac{m\pi}{h\beta^2}cos\left(\dfrac{m\pi z}{h} \right) \nabla_{t} \mathbf{Y}(\rho,\varphi)\\
\quad &=\dfrac{m\pi}{h\beta^2}cos\left(\dfrac{m\pi z}{h} \right) \hat{z} \times \left[ \hat{\rho}\dfrac{\partial}{\partial\rho} + \dfrac{\hat{\varphi}}{\rho}\dfrac{\partial}{\partial\varphi}\right]Y(\rho,\varphi) \\
\end{split}
\end{equation}
where $Y(\rho,\varphi)$ is now defines as:
\begin{equation}
Y(\rho, \varphi)= H_{0}J_{n}\left(\dfrac{\alpha_{nl}}{a} \rho\right)cos(n\varphi)
\label{angularTE}
\end{equation}

Then, the derivatives of  $ Y(\rho,\varphi) $ give the following results:
\begin{equation}
\begin{split}
&\dfrac{\partial Y(\rho,\varphi)}{\partial\rho}=\dfrac{\alpha^{'}_{nl}}{a}H_{0}J^{'}_{n}\left(\dfrac{\alpha^{'}_{nl}}{a} \rho\right)cos(n\varphi)\\ 
\quad &\dfrac{\partial Y(\rho,\varphi)}{\partial\varphi}=-nH_{0}J_{n}\left(\dfrac{\alpha^{'}_{nl}}{a} \rho\right)sin(n\varphi)
\end{split}
\end{equation}

Recalling that the first derivative of the Bessel function is given by equation \ref{BessDeriv}, the all set of the $TE_{nlm}$ modes are described by:
\begin{equation}
\begin{split}
H_{z}&=H_{0}J_{n}\left( \dfrac{\alpha^{'}_{nl}}{a} \rho\right) sin\left( \dfrac{m\pi z}{h}\right) cos\left( n \varphi\right) e^{\pm i \omega_{nlm} t}\\
H_{\rho}&=-\dfrac{m\pi a}{h \alpha^{'}_{nl}}H_{0}
\left[
\left(\dfrac{na}{\rho\alpha^{'}_{nl}}\right)
J_{n}\left(\dfrac{\alpha_{nl}}{a} \rho\right)- J_{n+1}\left(\dfrac{\alpha_{nl}}{a} \rho\right)
\right]\\
&\quad\quad\quad\quad\quad\quad\cdot cos\left( \dfrac{m\pi z}{h}\right)cos\left( n \varphi\right)  e^{\pm i \omega_{nlm} t}\\
\quad 
H_{\varphi}&=\dfrac{m n\pi a^2}{\rho h \alpha_{nl}^{2}}H_{0}J_{n}\left(\dfrac{\alpha^{'}_{nl}}{a}  \rho\right) cos\left( \dfrac{m\pi z}{h}\right) sin\left( n \varphi\right)  e^{\pm i \omega_{nlm} t}\\
\quad 
E_{z}&=0\\
\quad 
E_{\rho}&=\dfrac{i\mu_{0}\omega_{nlm} n a^2}{\rho \left(\alpha^{'}_{nl}\right)^{2}}H_{0}J_{n}\left(\dfrac{\alpha^{'}_{nl}}{a} \rho\right) sin\left( \dfrac{m\pi z}{h}\right) sin\left( n \varphi\right)  e^{\pm i \omega_{nlm} t}\\
\quad 
E_{\varphi}&=-\dfrac{i\mu_{0}\omega_{nlm} a}{\alpha^{'}_{nl}}H_{0}
\left[
\left(\dfrac{na}{\rho\alpha^{'}_{nl}}\right)
J_{n}\left(\dfrac{\alpha^{'}_{nl}}{a} \rho\right)- J_{n+1}\left(\dfrac{\alpha^{'}_{nl}}{a} \rho\right)
\right]\\
&\quad\quad\quad\quad\quad\quad\cdot sin\left( \dfrac{m\pi z}{h}\right)cos\left( n \varphi\right)  e^{\pm i \omega_{nlm} t}\\
\end{split}
\end{equation}
these modes are usually classified by the nomenclature $TE_{nlm}$, where the integers $n$, $l$ and $m$ measures the number of the sign changes of the field along its respective direction ($\rho$, $\varphi$ and $z$), as in the case of the $TM$ modes. Table \ref{ZeroBTab} shows some zeros of the first derivatives of the Bessel function.
\begin{table}
\centering
\caption{Zeros of the $n$-th Bessel function's first derivative.}
\begin{tabular}[c]{ c c c c c }
\hline
$l$ & $J^{'}_{0}(x)$ & $J^{'}_{1}(x)$ & $J^{'}_{2}(x)$ & $J^{'}_{3}(x)$ \\
\hline
1   &   3.8317   &   1.8412   &   3.0542   &   4.2012 \\
2   &   7.0156   &   5.3314   &   6.7061   &   8.0152 \\
3   &   10.1735  &   8.5363   &   9.9695   &   11.3459\\ 
4   &   13.3237  &   11.7060  &   13.1704  &   14.5858 \\
\hline
\end{tabular} 
\label{ZeroBTab}		
\end{table}

The fundamental resonating mode frequency is given by:
\begin{equation}
\omega_{111}=c\sqrt{\left(\dfrac{1.8412}{a}\right)^2+\dfrac{\pi^2}{h^2}}
\end{equation}
and the fields are:
\begin{equation}
\begin{split}
H_{z}&=H_{0}J_{1}\left( \dfrac{1.8412}{a} \rho\right) sin\left( \dfrac{\pi z}{h}\right) cos\left(\varphi\right) e^{\pm i \omega_{111} t}\\
H_{\rho}&=-\dfrac{\pi a}{1.8412 h }H_{0}
\left[
\left(\dfrac{a}{1.8412\rho }\right)
J_{1}\left(\dfrac{1.8412}{a} \rho\right)- J_{2}\left(\dfrac{1.8412}{a} \rho\right)
\right]\\
&\quad\quad\quad\quad\quad\quad\cdot cos\left( \dfrac{\pi z}{h}\right)  cos\left( \varphi\right) e^{\pm i \omega_{111} t}\\
\quad 
H_{\varphi}&=\dfrac{\pi a^2}{1.8412^{2}\rho h }H_{0}J_{1}\left(\dfrac{1.8412}{a}  \rho\right) cos\left( \dfrac{\pi z}{h}\right) sin\left(\varphi\right)  e^{\pm i \omega_{111} t}\\
\quad 
E_{\rho}&=\dfrac{i\mu_{0}\omega_{111} a^2}{1.8412^{2}\rho }H_{0}J_{1}\left(\dfrac{1.8412}{a} \rho\right) sin\left( \dfrac{\pi z}{h}\right) sin\left(\varphi\right)  e^{\pm i \omega_{111} t}\\
\quad 
E_{\varphi}&=-\dfrac{i\mu_{0}\omega_{111} a}{1.8412}H_{0}
\left[
\left(\dfrac{a}{1.8412\rho }\right)
J_{1}\left(\dfrac{1.8412}{a} \rho\right)- J_{2}\left(\dfrac{1.8412}{a} \rho\right)
\right]\\
&\quad\quad\quad\quad\quad\quad\cdot sin\left( \dfrac{\pi z}{h}\right) cos\left(\varphi\right)e^{\pm i \omega_{111} t}\\
\end{split}
\end{equation}
It is possible to notice that in this case the resonance frequency depends on both the length $h$ and the radius $a$ of the cavity. Indeed, the fields now vary along both  $\rho$ and $z$ directions.

\section{Conclusions}
In this paper the complete derivation of the $TM$ and $TE$ resonant modes in a cylindrical-shaped accelerating cavity from prime principles was presented.

The eigenvalues-eigenfunctions problem was approached with the separation of variables method, which allowed us to solve the wave equation for the system, introducing the three indexes $n$, $l$ and $m$ that fully describe the shape of the fields inside the resonator.

The role of the indexes $n$, $l$ and $m$ is to quantize the field oscillations in the cavity respect the radius $\rho$, the angle $\varphi$ and the length $z$ respectively. In particular, the value of these indexes gives an idea on the number of time that the field $\mathbf{E}$ or $\mathbf{B}$ changes sing in the direction quantized by that the particular index.

The most interesting class of modes of a cylindrical-shaped cavity is the $TM$ class. Indeed, the $TM_{0lm}$ modes are suitable to accelerate particles since they always possess a finite value of electric field at the cavity axis ($\rho=0$). The mode that is used in current particles colliders in $TM$ class cavities is the $TM_{010}$, which correspond the fundamental $TM$ mode.

The $TM_{1lm}$ modes are suitable to provide a transverse deflection to particle bunches. For these particular modes there is always a finite value of magnetic field at the cavity axis (but no $\mathbf{E}$) that is used to impose a transverse momentum to the traveling bunch.

\end{document}